\RequirePackage{cmap}
\documentclass{psta}
\usepackage{makecell}
\newcolumntype{P}[1]{>{\raggedright\arraybackslash\hspace{0pt}}p{#1}}
\newcolumntype{C}[1]{>{\centering\arraybackslash\hspace{0pt}}p{#1}}
\newcommand\thinhline{\Xhline{0.1pt}}

\def\colonTitlefont{\fontencoding{T2A}\fontfamily{cmr}\fontseries{m}\fontshape{n}\fontsize{6}{7}\selectfont}

\subjclass[UDC]{616.5-006:004.62:004.932}
\JournalRubric{2}
\ArticleType{RAR}

\title[Методика формирования набора дерматоскопических изображений]%
{Методика формирования клинически верифицированного набора дерматоскопических изображений}

\author{Козачок, {Е}лена {С}ергеевна}
\address{\ISPRAS}
\email{e.kozachok@ispras.ru}
\correspondent
\info{специалист ФГБУН «Институт системного программирования им. В.\,П.~Иванникова РАН» (ИСП~РАН). Научные интересы: медицинская информатика, машинное обучение в анализе медицинских изображений, дерматоскопия, формирование клинических наборов данных для задач компьютерного зрения.}
\orcid{0009-0007-9432-1663}

\begin{abstract}
Предмет исследования\,---\,методика формирования клинически верифицированного набора данных дерматоскопических изображений для задач медицинской информатики. Актуальность определяется тем, что качество автоматизированных систем диагностической поддержки зависит не только от объема изображений, но и от воспроизводимости процедуры съемки, полноты структурированных метаданных и надежности диагностических меток; международные коллекции при этом формировались в условиях, отличающихся от российской амбулаторной практики и мобильной дерматоскопии. Метод исследования\,---\,разработка трехкомпонентной методики, объединяющей стандартную операционную процедуру (СОП) получения изображения средствами мобильной дерматоскопии, информационную модель метаданных из 16 структурированных полей по шести клинически ориентированным блокам в нотации, совместимой с ISIC, и многоэтапную экспертную верификацию диагностической метки (первичная клиническая разметка, консенсусный пересмотр тремя специалистами, гистологическое подтверждение злокачественных новообразований). По методике в период с июня 2025~г. по май 2026~г. сформирован набор из 1026 уникальных дерматоскопических изображений от 443 пациентов; из 1044 исходных записей исключено 18 дублирующихся записей. Набор включает 9 нозологических категорий, все 39 изображений злокачественных классов (меланома, базалиома, плоскоклеточный рак) гистологически верифицированы; возраст пациентов 2--90 лет (медиана~38), в выборке 279 женщин и 164 мужчины. Каждое изображение сопровождается экспертно описанными дерматоскопическими структурами и явным индикатором уровня подтверждения диагноза \texttt{verification\_\allowbreak{}stage}, что позволяет использовать набор как пилотный клинически верифицированный ресурс для независимой оценки, анализа доменного сдвига, интерпретируемости моделей и последующего расширения. Практическая значимость\,---\,формализация воспроизводимого подхода к подготовке клинических дерматоскопических наборов, применимого при создании пилотных ресурсов для внешней проверки и методической демонстрации алгоритмов поддержки принятия врачебных решений в дерматоонкологии.
\end{abstract}

\title[Methodology for Creating a Dermoscopic Image Dataset]%
{Methodology for Creating a Clinically Verified Dermoscopic Image Dataset}

\author{Kozachok, {E}lena {S}ergeevna}
\address{\ISPRAS}
\email{e.kozachok@ispras.ru}
\info{specialist at the Ivannikov Institute for System Programming of the Russian Academy of Sciences. Research interests: medical informatics, machine learning for medical image analysis, dermoscopy, construction of clinical datasets for computer vision tasks.}
\orcid{0009-0007-9432-1663}

\begin{abstract}
This study presents a methodology for constructing a clinically verified dataset of dermatoscopic images for medical informatics research. The relevance of the work is driven by the fact that the performance of automated diagnostic support systems depends not only on the volume of images, but also on the reproducibility of the image acquisition procedure, the completeness of structured metadata, and the reliability of diagnostic labels. International collections were primarily created under conditions that differ substantially from routine Russian outpatient practice and mobile dermatoscopy. The proposed methodology integrates three interconnected components: (1)~a standard operating procedure (SOP) for acquiring images via mobile dermatoscopy, (2)~an information model comprising 16 structured metadata fields organized into six clinically oriented blocks in ISIC-compatible notation, and (3)~a multi-stage expert verification of diagnostic labels (initial clinical annotation, consensus review by three specialists, and histological confirmation of all malignant neoplasms). Using this methodology, a dataset of 1\,026 unique dermatoscopic images from 443 patients was collected between June~2025 and May~2026. From 1\,044 initial records, 18 duplicates were excluded. The dataset includes nine nosological categories; all 39 malignant lesions (18 melanomas, 15 basal cell carcinomas, and 6 squamous cell carcinomas) were histologically verified. Patient age ranged from 2 to 90 years (median~38), with 279 females (63\,\%) and 164 males (37\,\%). Each image is accompanied by expert-annotated dermatoscopic structures and an explicit \texttt{verification\_\allowbreak{}stage} field indicating the level of diagnostic confirmation. The resulting dataset serves as a pilot clinically verified resource suitable for independent model evaluation, domain shift analysis, interpretability studies, and further expansion.
\end{abstract}

\begin{document}
\English
\maketitle

\section*{Introduction}\label{sec:intro}

Dermoscopy is one of the primary tools for visual diagnosis of pigmented and non-pigmented skin neoplasms. The accuracy of dermoscopic image interpretation depends on the specialist's qualifications, image acquisition conditions, the completeness of the patient's clinical context, and the availability of morphological verification of the diagnosis. The development of deep learning methods shifts the requirements for diagnostic collections of dermoscopic images from the plane of visual material to the plane of structured information resources: for reproducible model training and independent evaluation, not only the images themselves are necessary, but also a consistent metadata information model, a documented acquisition protocol, and a transparent chain of diagnostic label verification.

International dermoscopic collections---HAM10000, ISIC, BCN20000, PAD-UFES-20---have played a decisive role in the development of automated skin lesion classification tasks \cite{ham10000,isic2017,isic2020,bcn20000,padufes20}. Accumulated experience has also shown that the transferability of models trained on such collections to local clinical environments is limited by domain shifts: differences in skin phototype distributions, image acquisition parameters, nosological referral structure, and the completeness of accompanying metadata \cite{daneshjou2022,groh2021,kinyanjui2020,esteva2017,wen2022}. For Russian outpatient practice, where dermoscopy is performed primarily using mobile devices (optical dermatoscope combined with a smartphone), these limitations are fully manifested, which justifies the development of locally verified datasets.

The key methodological challenge is that a dermoscopic dataset is not a simple collection of image files. To be suitable for medical informatics tasks, three problems must be solved simultaneously: ensuring reproducibility of image acquisition conditions, agreeing on a structured metadata composition, and recording the reliability level of each diagnostic label. Each of these components is addressed separately in the literature on open datasets \cite{isic2020,padufes20,brinker2019}, but in medical informatics literature, a unified methodology in which all three components are integrated and mutually consistent is rarely presented.

\textbf{Objective} --- to develop a methodology for constructing a clinically verified dermoscopic image dataset of skin neoplasms, integrating a standard operating procedure (SOP) for image acquisition, a metadata information model, and multi-stage expert verification of diagnostic labels, and to construct a dataset suitable for medical informatics tasks using this methodology.

\textbf{Research objectives:} formulate requirements for the information model of a clinical dermoscopic dataset; develop an SOP for image acquisition using mobile dermoscopy; define a metadata structure consistent with international dermoscopic collection practices; formalize a multi-stage expert verification procedure; construct a dataset using the methodology and provide its statistical description; identify limitations and directions for development.

\textbf{Scientific novelty} of the work:
\begin{enumerate}
\item A three-component methodology for constructing a clinical dermoscopic dataset has been formulated, in which the acquisition SOP, the metadata information model, and multi-stage diagnostic verification are treated as interrelated elements of a single reproducible information resource; such an integrated formulation has not previously been presented in the literature on open dermoscopic collections.
\item A metadata information model has been developed, organized into six clinically oriented blocks and consistent with international dermoscopic collection practices in field notation, while including a field for expert annotation of dermoscopic structures, which creates the basis for quantitative comparison of model attention areas with clinically significant image zones.
\item A multi-stage procedure for expert verification of diagnostic labels has been formalized with explicit recording of the reliability level in each record (initial annotation --- consensus of three specialists --- histological confirmation of malignant neoplasms), enabling the dataset to be used as a pilot resource for independent evaluation, domain shift analysis, and interpretability studies accounting for label reliability heterogeneity.
\item A clinically verified dataset of 1026 images from 443 patients collected in Russian outpatient practice has been constructed, covering 9 nosological categories and supported by histological verification of malignant classes.
\end{enumerate}

\textbf{Practical significance} consists in the reproducibility of the approach to preparing a clinical dermoscopic dataset: the proposed methodology can be applied when constructing new medical image collections, independently testing classification algorithms, developing clinical decision support systems in dermato-oncology, and comparatively evaluating models under conditions close to Russian outpatient practice.

\section{Review of Existing Dermoscopic Image Datasets}\label{sec:review}

A comparison of existing dermoscopic datasets is more meaningful with respect to information model composition and diagnostic verification characteristics than by volume alone, as these parameters determine the dataset's suitability for reproducible model training and independent evaluation. The classic reference dataset is PH2~\cite{ph2}---200 dermoscopic images in three clinical categories (common nevus, atypical nevus, melanoma) from a single center with histological confirmation; limited size is compensated by homogeneous acquisition conditions. HAM10000~\cite{ham10000} contains 10\,015 images across seven diagnostic categories from two clinical centers and remains one of the most frequently cited benchmarks. The ISIC~2017 challenge~\cite{isic2017} was one of the first open platforms for reproducible model evaluation, and the ISIC~2019 collection, combining HAM10000, BCN, and part of MSK, expanded the training set to 25\,331 images across eight classes~\cite{combalia2022}. ISIC~2020~\cite{isic2020} (33\,126 images from 2\,056 patients, full histological verification) was the first to formally document the acquisition procedure and expand clinical metadata. BCN20000~\cite{bcn20000} describes 18\,946 images from a single large dermatology center and records extensive patient clinical context. PAD-UFES-20~\cite{padufes20} was collected using consumer-grade cameras and is accompanied by a clinical feature questionnaire, but formally represents clinical rather than dermoscopic images of skin lesions.

Systematic reviews of publicly available dermoscopic image datasets~\cite{wen2022} and a separate review on the lack of transparency in dermatology AI datasets and algorithms~\cite{daneshjou2021} identify several persistent limitations: opaque diagnosis confirmation procedures, limited skin phototype information, absence of documented acquisition SOPs, and non-uniform metadata fields across collections. The concept of ``datasheets for datasets''~\cite{gebru2021} formalizes requirements for describing sources, collection procedures, metadata composition, and sampling limitations, and is now considered one of the standard documentation tools in medical machine learning. Studies on biases and representativeness of dermatological algorithms~\cite{daneshjou2022,groh2021,kinyanjui2020} have shown that insufficient representation of certain phototypes and population groups leads to reduced model robustness when applied in environments differing from the training set. Deep learning approaches to differential diagnosis of melanoma and skin cancer achieve expert-level performance only when training and target distributions are sufficiently aligned~\cite{brinker2019,winkler2019}; otherwise, systematic biases arise related to sample composition, data source, and acquisition conditions~\cite{winkler2019}.

A comparative characterization of international datasets and the developed dataset is provided in Table~\ref{tab:datasets}. The table is not intended to rank datasets: its purpose is to show which properties should be explicitly documented when preparing a clinical dermoscopic dataset for medical informatics tasks.

\begin{table}\scriptsize
\caption{Comparative characterization of dermoscopic image datasets}\label{tab:datasets}
\centering
\setlength{\tabcolsep}{1pt}
\renewcommand{\arraystretch}{1.15}
\begin{tabular}{|P{20mm}|C{7mm}|C{15mm}|C{10mm}|P{15mm}|P{13mm}|P{18mm}|P{10mm}|}
\hline
Dataset & Year & Images & Class. & Acquisition & Metadata & Diagnosis confirmation & SOP\\
\hline
PH2~\cite{ph2} & 2013 & 200 & 3 & Single-center dermoscopy & Minimal clinical features & Histology & Not standardized\\
\thinhline
HAM10000~\cite{ham10000} & 2018 & 10\,015 & 7 & Dermoscopy, 2 centers & Age, sex, site, diagnosis & Histology, clinical, or consensus & Not described\\
\thinhline
ISIC~2017~\cite{isic2017} & 2018 & 2\,750 & 3 & Dermoscopy, multi-source & Minimal & Mixed modes & Not standardized\\
\thinhline
ISIC~2019~\cite{combalia2022} & 2019 & 25\,331 & 8 & HAM10000 + BCN + MSK & Depends on subcollection & Mixed modes & Not standardized\\
\thinhline
ISIC~2020~\cite{isic2020} & 2020 & 33\,126 (2\,056~pat.) & 9{+}bin. & Multi-center dermoscopy & Extended clinical + context & Histology & Documented\\
\thinhline
BCN20000~\cite{bcn20000} & 2024 & 18\,946 & 8 & Single large center dermoscopy & Clinical and admin. data & Expert + morphology (partial) & Partially described\\
\thinhline
PAD-UFES-20~\cite{padufes20} & 2020 & 2\,298 (1\,373~pat.) & 6 & Smartphone, clinical & Patient and lesion questionnaire & Partial histology + clinical & Described for smartphone\\
\thinhline
\textbf{This study} & \textbf{2026} & \textbf{1\,026 (443~pat.)} & \textbf{9} & \textbf{Mobile dermoscopy} & \textbf{16 fields, ISIC-compatible} & \textbf{Multi-stage: annotation $\to$ consensus of~3 $\to$ histol.\ malignant} & \textbf{Documented}\\
\hline
\end{tabular}
\end{table}

From the comparison, it follows that the tasks of a clinical dermoscopic dataset construction methodology include, at minimum, documented description of image acquisition conditions, a structured metadata information model, and a transparent diagnosis confirmation procedure. These three elements form the basis of the proposed methodology. The developed dataset is substantially smaller than ISIC~2019/2020 or BCN20000, but surpasses comparable-sized datasets (PH2, PAD-UFES-20) in structured metadata completeness and formalization of the diagnostic label verification chain.

\section{Materials and Methods}\label{sec:methods}

\subsection{Methodology Architecture}

\begin{figure}
\centering
\includegraphics[width=\linewidth]{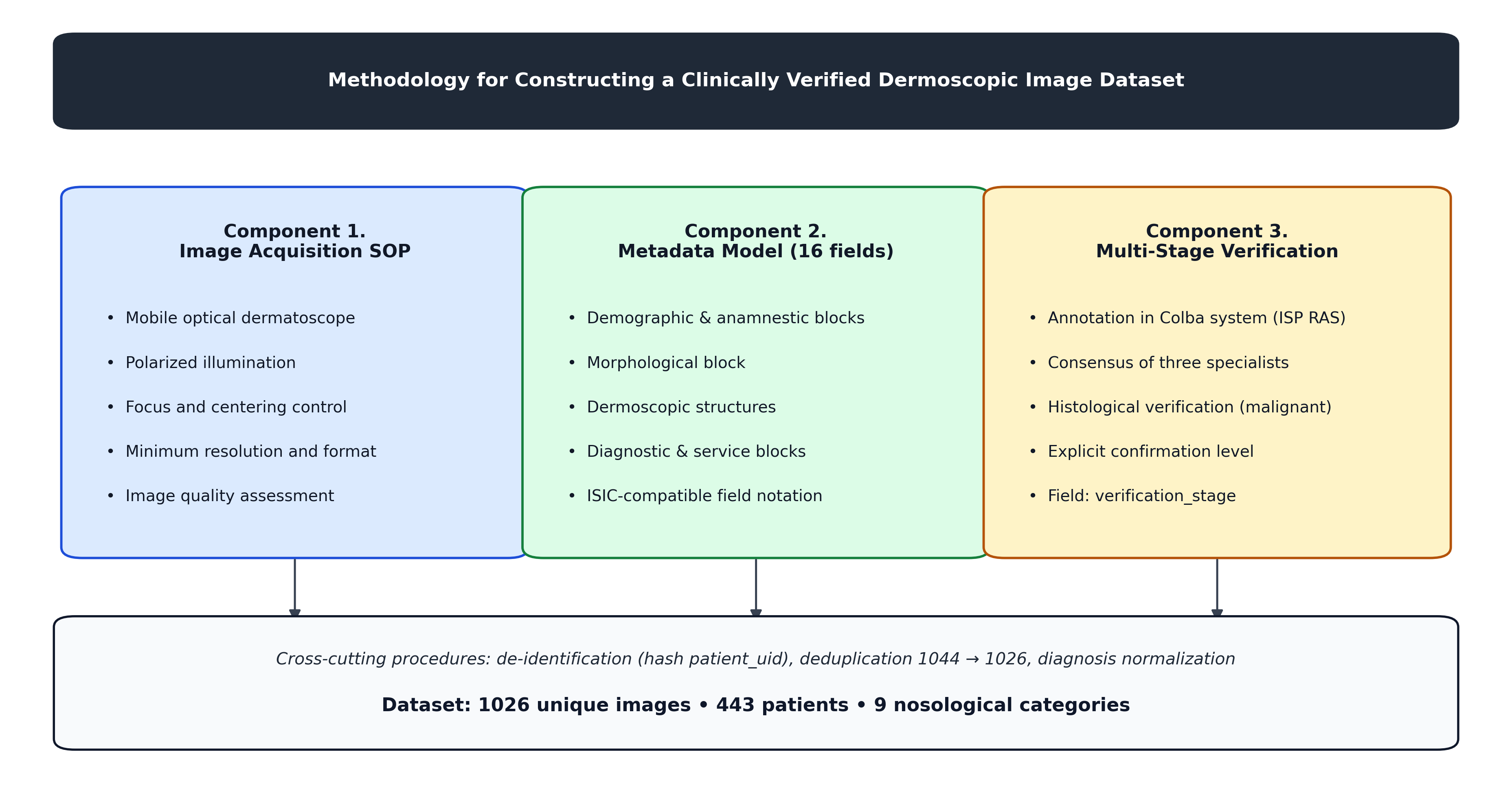}
\caption{Architecture of the methodology for constructing a clinically verified dermoscopic image dataset}\label{fig:scheme}
\end{figure}

The developed methodology treats the construction of a clinical dermoscopic dataset as a reproducible information process in which each image receives an interpretable description and a label with an explicit reliability level. The methodology is structurally composed of three interrelated components: a standard operating procedure for image acquisition, a metadata information model, and multi-stage expert verification of diagnostic labels. Cross-cutting procedures include patient record de-identification, deduplication, and normalization of diagnostic categories. The overall methodology architecture is presented in Figure~\ref{fig:scheme}.

The processing object of the methodology is an image--patient pair. Input consists of a dermoscopic image and the associated clinical record; output is a de-identified dataset record containing the image, diagnostic label, structured metadata, and a verification level attribute. Quality control is embedded in each component: at the acquisition stage---through technical image quality assessment; at the description stage---through a formal metadata schema; and at the verification stage---through explicit accounting for the diagnosis confirmation level.

\subsection{Period and Data Sources}

The dataset was constructed prospectively from June 2025 to May 2026. The image recording unit is a unique dermoscopic image file; the patient recording unit is a de-identified patient record. Included were dermoscopic images of skin neoplasms acquired in accordance with the developed SOP and accompanied by a clinical conclusion and a minimum set of structured metadata. Multiple images per patient were permitted, since in outpatient practice a single patient may have several neoplasms requiring separate evaluation. Excluded from processing were records without an image file, duplicates identified by file content hash, and records with values of key fields unsuitable for statistical analysis.

\subsection{Ethical Aspects of Data Management}

The study was conducted in accordance with the principles of the Declaration of Helsinki. The collection of dermoscopic images and clinical information was carried out within the framework of routine outpatient dermoscopic practice; written informed consent was obtained from each patient for the use of de-identified clinical data and dermoscopic images for research purposes.

De-identification of records was performed prior to inclusion in the dataset: the personal patient identifier was replaced with a stable de-identified code; images contain no patient-identifying features (face, tattoos, or other unique skin markers in the field of view are absent or excluded during cropping).

The dataset is non-public and is used exclusively within the closed research environment of the Ivannikov Institute for System Programming of the Russian Academy of Sciences. Publication of images in open access is not planned.

\subsection{Component 1. Standard Operating Procedure for Image Acquisition}\label{sec:sop}

The standard operating procedure (SOP) was developed for mobile dermoscopy conditions and specifies requirements for equipment, acquisition conditions, and technical quality control of each image. Image acquisition is performed using an optical dermatoscope coupled with a smartphone; this configuration corresponds to outpatient practice and allows acquisition at various clinical sites without specialized stationary equipment. This configuration is associated in the literature with mobile dermoscopy and described as a distinct class of imaging devices~\cite{padufes20,brinker2019}.

Basic SOP requirements are organized into four groups:
\begin{enumerate}
\item \textbf{Optical conditions.} Polarized dermoscopy is used in contact or immersion mode; dermatoscope tilt angles relative to the skin surface and contact pressure are standardized to avoid compression artifacts.
\item \textbf{Acquisition parameters.} Framing is performed so that the neoplasm is centered in the field of view and occupies the main portion of the frame; focus is checked before saving the image; minimum resolution and file format are regulated.
\item \textbf{Illumination conditions.} The dermatoscope's built-in polarized light source is used; mixed illumination that alters image color balance is prohibited.
\item \textbf{Technical quality control.} Each image is assessed using a composite quality score \texttt{image\_quality\_score} on an ordinal scale from 1 to 5: 1---image is unsuitable (blurring, substantial glare, extensive obscuration of the neoplasm by artifacts); 2---significant defects, but the object of interest is discernible; 3---acceptable quality with local artifacts; 4---good quality, minimal artifacts; 5---reference quality. Images scoring 1--2 are excluded by the specialist directly at the acquisition stage and do not proceed to further processing; images scoring 3--5 are included in the array. The \texttt{image\_quality\_score} value is stored in metadata for subsequent stratification. During the dataset construction period, 32 images were excluded at the SOP stage based on \texttt{image\_quality\_score} $\leq 2$.
\end{enumerate}

The SOP also specifies the specialist's workflow: initial visual inspection of the skin, selection of the acquisition target, skin preparation (if necessary), acquisition of multiple frames, and selection of the representative image. Documentation of actions is performed in a structured form, ensuring reproducibility across different clinical sites and consistency between specialists.

\subsection{Component 2. Metadata Information Model}\label{sec:metadata}

Metadata are organized not as an arbitrary list of fields, but as a clinically oriented information model consisting of six blocks. The model composition is presented in Table~\ref{tab:metadata}. Field structure and notation are consistent with international dermoscopic collection practices, facilitating joint processing with open sources and independent model evaluation.

\begin{table}\footnotesize
\caption{Metadata information model of the developed dataset (16~fields)}\label{tab:metadata}
\centering
\setlength{\tabcolsep}{3pt}
\renewcommand{\arraystretch}{1.2}
\begin{tabular}{|P{20mm}|P{50mm}|P{30mm}|}
\hline
Block & Fields & Purpose\\
\hline
Demographic & \texttt{age\_at\_exam}, \texttt{sex}, \texttt{fitzpatrick\_type} & Basic patient characteristics and skin phototype (Fitzpatrick I--IV)\\
\thinhline
Anamnestic & \texttt{sunburn\_history}, \texttt{personal\_ca\_history} & Risk factors: sunburn history, oncological history\\
\thinhline
Morphological & \texttt{anatomical\_site}, \texttt{lesion\_diameter\_mm}, \texttt{dominant\_colors}, \texttt{elevation}, \texttt{border\_regularity} & Site, size, color, and geometric features of the neoplasm\\
\thinhline
Dermoscopic & \texttt{dermoscopic\_\allowbreak{}structures} & Expert annotation of patterns (network, globules, pseudopods, blue veil, vascular structures)\\
\thinhline
Diagnostic & \texttt{clinical\_diagnosis}, \texttt{histopathology\_result}, \texttt{verification\_\allowbreak{}stage} & Clinical diagnosis, histological result, verification stage\\
\thinhline
Service & \texttt{examiner\_id}, \texttt{image\_quality\_score} & De-identified specialist identifier and technical image quality score\\
\hline
\end{tabular}
\end{table}

The dermoscopic block is highlighted separately, as the \texttt{dermoscopic\_\allowbreak{}structures} field has independent methodological significance. Expert-annotated patterns represent clinically interpretable features that physicians rely on for differential diagnosis. Recording them enables the dataset to serve not only as a source of classification labels, but also as a resource for quantitative comparison of machine learning model attention areas with clinically significant image zones~\cite{combalia2022,tschandl2020}, creating the basis for subsequent interpretability studies in dermato-oncology.

The link between multiple images and a single patient is maintained through a stable de-identified identifier. Original personal data are not included in the dataset; the record structure nonetheless enables patient-stratified splitting and repeated-visit analysis.

\subsection{Component 3. Multi-Stage Expert Verification}\label{sec:verif}

Diagnostic verification is structured as a multi-stage procedure that enhances the reliability of the diagnostic label and records its level in the explicit \texttt{verification\_\allowbreak{}stage} field. The stages and their purposes are presented in Table~\ref{tab:verif}.

\begin{table}\footnotesize
\caption{Stages of expert verification of diagnostic labels}\label{tab:verif}
\centering
\setlength{\tabcolsep}{3pt}
\renewcommand{\arraystretch}{1.2}
\begin{tabular}{|P{27mm}|P{47mm}|P{32mm}|}
\hline
Stage & Content & Result\\
\hline
Initial clinical annotation & Dermatologist records clinical diagnosis and dermoscopic structure description according to the metadata model form & Initial label with specialist identifier\\
\thinhline
Consensus review & Independent review of the record by three specialists; disagreements resolved through discussion and a consensus formulation recorded & Confirmed or corrected label\\
\thinhline
Histological verification & For cases with suspected malignancy, the record is supplemented with histopathological examination results & Label with morphological confirmation\\
\hline
\end{tabular}
\end{table}

Initial annotation and consensus review are performed in the Colba image annotation system developed at the Ivannikov ISP RAS. The environment provides structured entry of information model fields, support for collaborative annotation by multiple specialists, maintenance of a diagnostic label change log, and export of de-identified records in a format compatible with subsequent software processing. Use of a specialized annotation environment, rather than general-purpose annotation tools, restricts entered values to a predefined nomenclature and reduces the risk of terminology inconsistency between specialists.

Inter-rater agreement at the independent review stage is treated as a separate quality control metric for annotation. For subsequent dataset versions that retain each expert's initial decisions prior to consensus discussion, it can be quantitatively assessed using a multi-rater agreement coefficient (e.g., Fleiss's $\kappa$) and included in the dataset datasheet.

This approach enables separation of the initial clinical hypothesis from the confirmed diagnostic label and preserves the reliability level of each record. Records at different verification levels are not discarded but labeled: the confirmation stage becomes a property of the record and can be explicitly used when selecting subsets for training, validation, and independent testing. Clinical research conventionally distinguishes between ``gold standard'' (histology) and ``clinical confirmation,'' and the proposed methodology implements this distinction directly in the dataset information model.

In the constructed dataset, all 1026 unique images underwent consensus review by three specialists and have \texttt{verification\_\allowbreak{}stage}~$\geq 2$. All 39 images of malignant categories (melanoma---18, basal cell carcinoma---15, squamous cell carcinoma---6) are additionally accompanied by histopathological examination results and have \texttt{verification\_\allowbreak{}stage}~$= 3$. Benign and borderline neoplasms retain \texttt{verification\_\allowbreak{}stage}~$= 2$, since biopsy is not clinically indicated for these cases. This scheme enhances the reliability of diagnostic labels for malignant classes and enables the dataset to be used for independent algorithm evaluation with explicit consideration of each record's verification level.

\subsection{Cross-Cutting Procedures: Deduplication and Diagnostic Category Normalization}

The initial array included 1044 records. Deduplication was performed using a cryptographic hash of image file content: a stable hash was computed for each original image, and records with matching values were treated as a single graphic object regardless of filename, container format, or re-uploaded data. After exclusion of 18 duplicate records, the final array comprised 1026 unique images; the proportion of excluded duplicates was 1.7\,\% of the initial record count. The hash is stored in the record's service fields and ensures reproducibility of the deduplication procedure in subsequent dataset maintenance.

Diagnostic categories were normalized to a unified nomenclature comprising 9 classes: nevus, dysplastic nevus, melanoma, seborrheic keratosis, hemangioma, dermatofibroma, basal cell carcinoma, squamous cell carcinoma, papilloma. Records with an uncertain service category of benign nevus-like formation were assigned to the nevus class in the absence of clinical signs of malignancy. For benign neoplasms, the final label was determined by clinical assessment and expert consensus; histological conclusion was considered only when available in the clinical record.

Nosological categories are described along two independent attributes (Table~\ref{tab:diag}): neoplasm origin (melanocytic or non-melanocytic) and clinical status (benign, borderline, or malignant). Dysplastic nevi are designated as a separate borderline group: morphologically they belong to melanocytic neoplasms and may exhibit atypical features requiring differential diagnosis with melanoma, but are not malignant by definition. In forming a 2-class ``benign/malignant'' task (e.g., for a screening scenario), dysplastic nevi are by default assigned to the benign group, since their referral for biopsy is determined clinically. The dataset user may change this policy and assign them to the malignant class or form a separate class---the \texttt{clinical\_diagnosis}, origin, and clinical status fields in Table~\ref{tab:diag} preserve the original nosological annotation without loss of information.

\subsection{Statistical Analysis}

Statistical description of the dataset included counts of images and patients, distributions by nosological category, sex, age, and number of images per patient. For age, minimum and maximum values, mean, and median were calculated. For categorical variables, absolute and relative frequencies were calculated. Calculations were performed on de-identified final dataset tables; graphs were constructed from aggregated quantitative data.

\section{Results}\label{sec:results}

\subsection{General Dataset Characteristics}

The application of the methodology yielded a dataset of 1026 unique dermoscopic images from 443 patients. Images per patient range from 1 to 10; median is 2 images, mean is 2.32 images; 190 patients contributed a single image. The dataset contains melanocytic and non-melanocytic neoplasms: melanocytic---767 images (74.8\,\%), non-melanocytic---259 images (25.2\,\%). Malignant neoplasms total 39 images (3.8\,\%): 18 melanomas, 15 basal cell carcinomas, and 6 squamous cell carcinomas.

\begin{table}\scriptsize
\caption{Distribution of images by nosological category}\label{tab:diag}
\centering
\setlength{\tabcolsep}{1pt}
\renewcommand{\arraystretch}{1.15}
\begin{tabular}{|P{32mm}|r|r|P{25mm}|P{23mm}|}
\hline
Nosological category & N & Share,\,\% & Origin & Clinical status\\
\hline
Nevus & 733 & 71.4 & Melanocytic & Benign\\
\thinhline
Seborrheic keratosis & 118 & 11.5 & Non-melanocytic & Benign\\
\thinhline
Hemangioma & 87 & 8.5 & Non-melanocytic & Benign\\
\thinhline
Dermatofibroma & 31 & 3.0 & Non-melanocytic & Benign\\
\thinhline
Melanoma & 18 & 1.8 & Melanocytic & Malignant\\
\thinhline
Dysplastic nevus & 16 & 1.6 & Melanocytic & Borderline\\
\thinhline
Basal cell carcinoma & 15 & 1.5 & Non-melanocytic & Malignant\\
\thinhline
Squamous cell carcinoma & 6 & 0.6 & Non-melanocytic & Malignant\\
\thinhline
Papilloma & 2 & 0.2 & Non-melanocytic & Benign\\
\hline
Total & 1026 & 100.0 & \,---\, & \,---\,\\
\hline
\end{tabular}
\end{table}

\begin{figure}
\centering
\includegraphics[width=\linewidth]{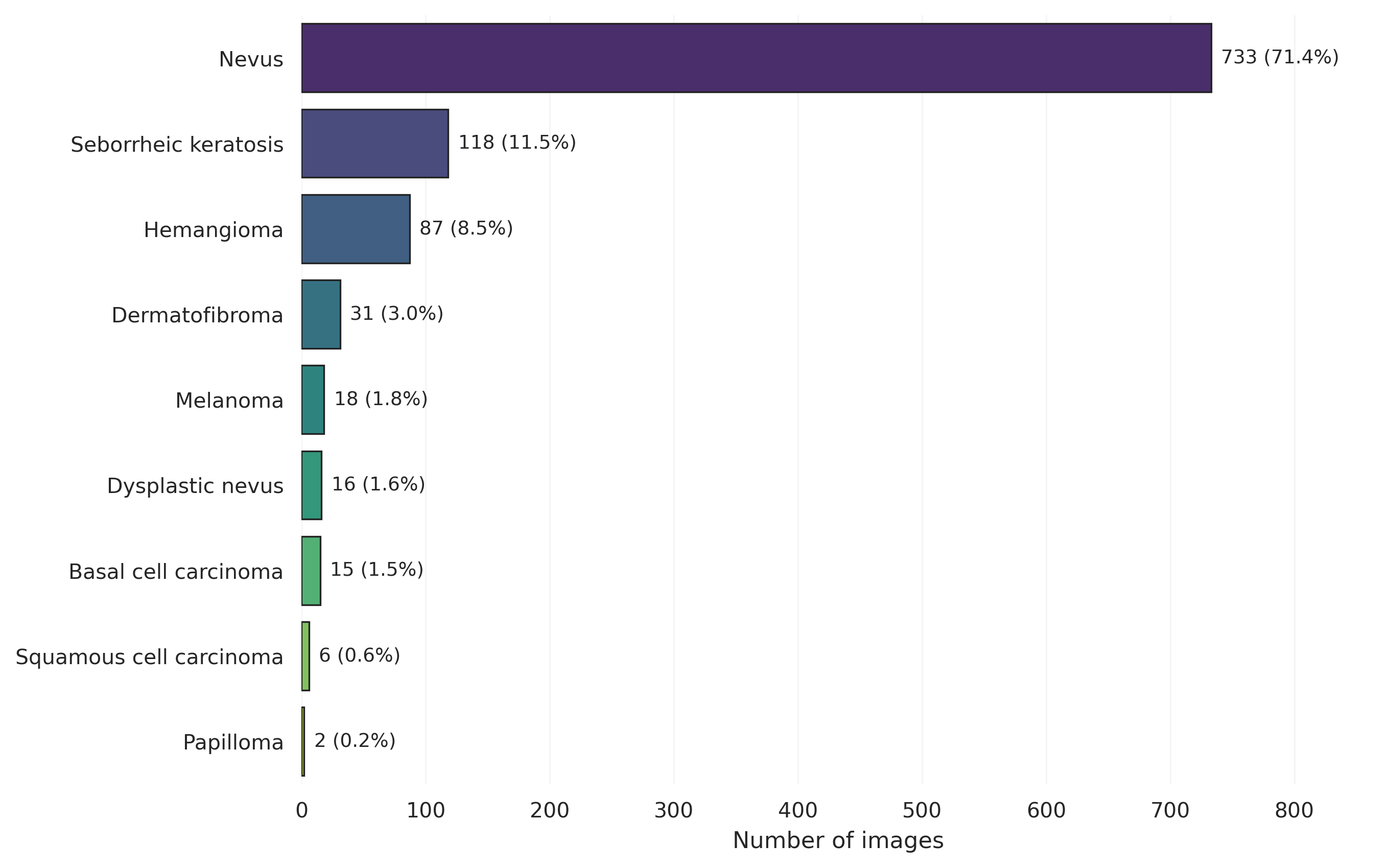}
\caption{Distribution of images by nosological category}\label{fig:diag}
\end{figure}

The predominance of nevi is expected for outpatient dermoscopic practice, since benign pigmented neoplasms are a frequent reason for preventive visits and require differential diagnosis with melanoma. At the same time, this distribution demonstrates a pronounced class imbalance that must be accounted for in training and evaluating machine learning algorithms. A graphical representation of the distribution is shown in Figure~\ref{fig:diag}.

\subsection{Demographic Structure}

\begin{figure}
\centering
\includegraphics[width=\linewidth]{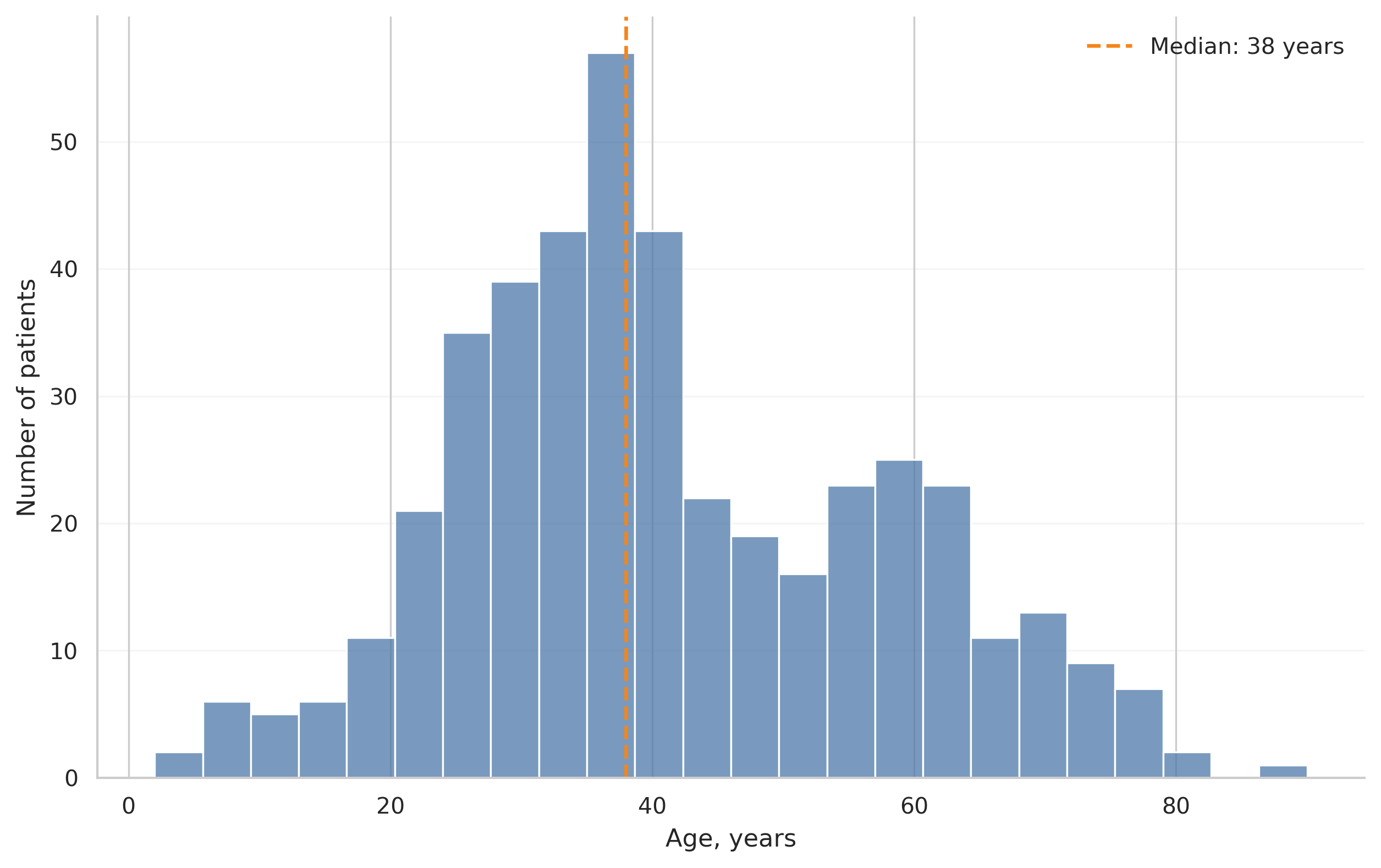}
\caption{Age distribution of patients}\label{fig:age}
\end{figure}

After exclusion of invalid age values, age analysis included 439 patients. Age ranged from 2 to 90 years; median was 38 years, mean was 41.1 years. The age distribution is shown in Figure~\ref{fig:age}.

\begin{figure}
\centering
\includegraphics[width=0.8\linewidth]{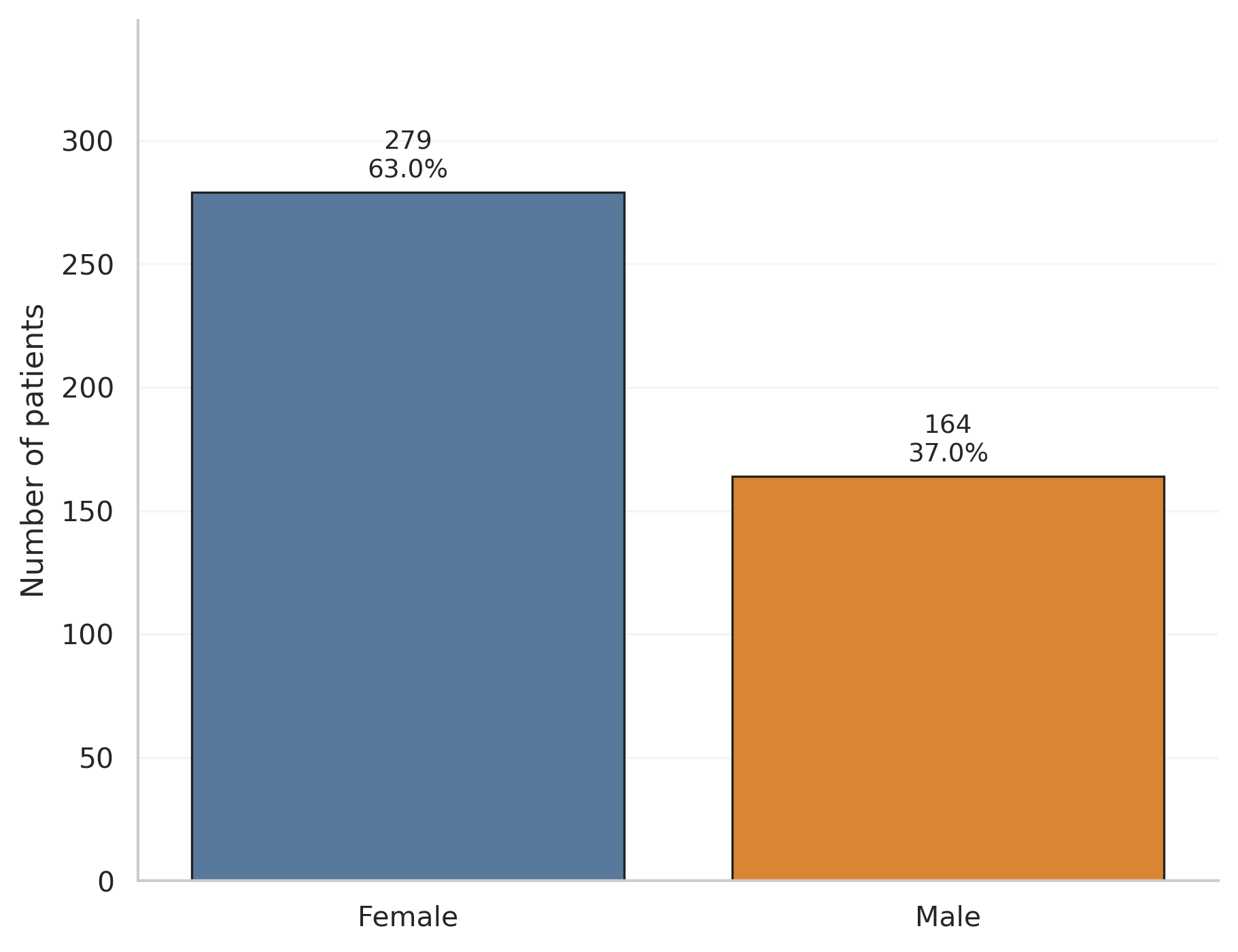}
\caption{Distribution of patients by sex}\label{fig:sex}
\end{figure}

The sample included 279 females (63.0\,\%) and 164 males (37.0\,\%). The predominance of females may be related to outpatient dermoscopic screening visit patterns and should not be interpreted as a population-level estimate of skin neoplasm prevalence. The sex distribution is shown in Figure~\ref{fig:sex}.

\subsection{Images per Patient}

\begin{figure}
\centering
\includegraphics[width=\linewidth]{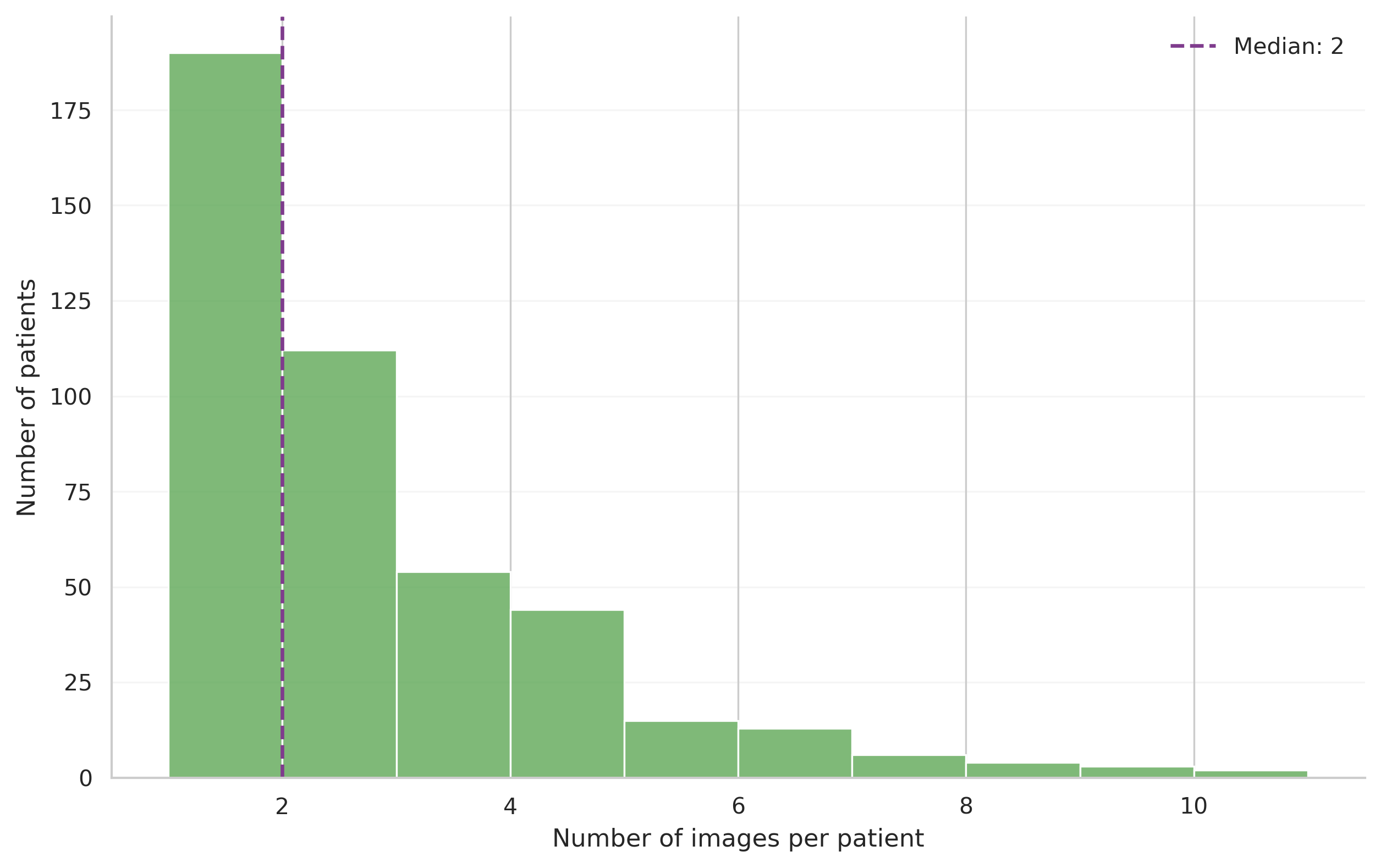}
\caption{Number of images per patient}\label{fig:imgs_per_patient}
\end{figure}

Since a single patient may have multiple skin neoplasms, the distribution of images per patient is a separate dataset characteristic (Figure~\ref{fig:imgs_per_patient}). The presence of multiple images for some patients reflects the real clinical scenario of a dermoscopic examination, in which the physician evaluates not one isolated neoplasm but several clinically significant lesions. This property has methodological significance when splitting the dataset into training and test sets: patient-stratified splitting is required to prevent overestimation of algorithm performance.

\subsection{Summary Dataset Characteristics}

Key quantitative characteristics of the dataset are presented in Table~\ref{tab:control}. They are calculated from de-identified final dataset tables and describe the completeness and consistency of the constructed array.

\begin{table}\footnotesize
\caption{Summary characteristics of the constructed dataset}\label{tab:control}
\centering
\setlength{\tabcolsep}{3pt}
\renewcommand{\arraystretch}{1.15}
\begin{tabular}{|P{78mm}|r|}
\hline
Indicator & Value\\
\hline
Excluded at SOP stage by \texttt{image\_quality\_score} $\leq 2$ & 32\\
\thinhline
Initial records (after SOP quality control) & 1\,044\\
\thinhline
Excluded duplicates (by cryptographic hash) & 18\\
\thinhline
Proportion of excluded duplicates, \% & 1.7\\
\thinhline
Unique images & 1\,026\\
\thinhline
Patients with images & 443\\
\thinhline
Min / median / mean / max images per patient & 1\,/\,2\,/\,2.32\,/\,10\\
\thinhline
Diagnostic categories & 9\\
\thinhline
Fields in the metadata information model & 16\\
\thinhline
Images with \texttt{verification\_\allowbreak{}stage}~$=2$ (consensus) & 987\\
\thinhline
Images with \texttt{verification\_\allowbreak{}stage}~$=3$ (histology) & 39\\
\thinhline
Proportion of histologically verified of all malignant, \% & 100.0\\
\hline
\end{tabular}
\end{table}

\section{Discussion}\label{sec:discussion}

The proposed methodology treats the construction of a clinical dermoscopic dataset as an independent medical informatics task. Its novelty consists not in individual technical techniques, but in the simultaneous formalization of three components---the image acquisition SOP, the metadata information model, and multi-stage expert verification---treated as interrelated elements of a single information resource. Compared to an approach in which a dataset is described only by its image count and class list, this structure records the provenance of each record, its link to a de-identified patient record, the structured metadata composition, and the diagnosis confirmation level. This is particularly important for subsequent deep learning applications, where annotation errors, duplicates, and incomplete clinical context can substantially distort model quality estimates~\cite{wen2022,brinker2019,winkler2019,ph2,combalia2022,daneshjou2021,gebru2021}.

The description of a standard image acquisition procedure translates acquisition requirements from ``clinical expertise'' to verifiable technological conditions. This is important for two reasons. First, the described SOP enhances reproducibility across multiple clinical sites and reduces variability in input images between specialists. Second, explicit regulation of mobile dermoscopy parameters (polarization, focus, centering, quality assessment) allows analytical separation of distortions related to acquisition conditions from distortions attributable to model properties. In international dermoscopic collections, such regulation is inconsistently represented~\cite{wen2022}, which complicates independent model comparison.

The 16-field metadata information model extends the dataset's applicability. Demographic and anamnestic blocks enable assessment of algorithm stability across patient subgroups; morphological and dermoscopic blocks enable analysis of clinical differential diagnosis factors; the diagnostic block enables correct separation of clinical hypothesis, morphological confirmation, and verification level. The \texttt{dermoscopic\_\allowbreak{}structures} field is of particular significance: it creates a quantitative basis for model interpretability analysis by enabling comparison of classifier high-activation areas with expert-annotated image patterns~\cite{combalia2022,tschandl2020}. The dataset thus creates conditions for subsequent explainable AI research in dermato-oncology.

Multi-stage expert verification forms a transparent chain of diagnostic label reliability. The \texttt{verification\_\allowbreak{}stage} value is preserved in the record and used as a full attribute when selecting subsets for training and testing. Histological verification of malignant classes corresponds to the ``gold standard'' concept in clinical diagnosis and substantially reduces the risk of systematic errors in classifier sensitivity evaluation for malignant neoplasms.

\subsection{Relationship to Previous Publications}

A pilot dataset on clinical dermoscopic dataset construction was previously published by the author~\cite{kozachok2025dataset}: it included 657 dermoscopic images with annotation of clinically significant features accounting for skin phototypes of the Russian Federation population. Applied research was performed on this material using the dataset as a data source rather than an object of methodological formalization: evaluation of multi-class and cascade two-stage classification algorithms on the proprietary dataset and the ISIC-2019 repository~\cite{kozachok2025screening}, architectural description of an intelligent clinical decision support system implementing the developed algorithms~\cite{kozachok2025sppvr1}, and clinical validation of a screening scenario using mobile dermoscopy with expert agreement assessment~\cite{kozachok2025sppvr2}.

The present work introduces a new clinically verified dataset constructed using a formalized methodology, rather than a continuation of the pilot dataset described earlier~\cite{kozachok2025dataset}. Compared to the preceding research phase, the data preparation principle has changed: (i)~the dataset was constructed anew and includes 1026 unique images from 443 patients with explicit recording of the image-per-patient distribution; (ii)~a standard operating procedure for mobile dermoscopic acquisition was formulated, regulating optical conditions, acquisition parameters, illumination conditions, and technical image quality control; (iii)~the metadata list was organized into an information model of 16 structured fields in six clinically oriented blocks in ISIC-compatible notation, with explicit designation of the expert dermoscopic structure annotation field; (iv)~the diagnostic label is determined through a multi-stage expert verification procedure (initial annotation $\to$ consensus of three specialists $\to$ histological confirmation of malignant neoplasms) with reliability level recorded in the service field \texttt{verification\_\allowbreak{}stage}; (v)~deduplication was performed using a cryptographic hash of image file content and removed 18 duplicate records from 1044 original records.

The present article therefore does not duplicate the previous description~\cite{kozachok2025dataset}, but presents an independent research resource and formalizes the three-component construction methodology, designating the acquisition SOP, metadata information model, and multi-stage verification as independent interrelated components. Applied research~\cite{kozachok2025screening,kozachok2025sppvr1,kozachok2025sppvr2} performed earlier on pilot material is treated as a preceding stage and does not determine the composition of the current dataset.

\smallskip
The applied cryptographic hash-based deduplication is robust to filename and container format differences and maintains procedural reproducibility through the record's service hash. Prospectively, the procedure can be extended with perceptual hashing to detect visually similar images that do not match byte-for-byte.

The developed dataset does not replace large open collections but complements them as a local clinical resource with an explicitly documented information model. Its value lies in the combination of mobile dermoscopy, extended patient and dermoscopic metadata, and multi-stage verification; such a resource can be used for independent testing of models trained on open sources and for investigating domain adaptation methods for Russian outpatient practice.

\section{Limitations}\label{sec:limitations}

The constructed dataset has several limitations that should be considered when using it:
\begin{enumerate}
\item \textbf{Class imbalance.} The distribution of nosological categories reflects outpatient referral patterns and is characterized by pronounced predominance of nevi; the small number of images in certain malignant categories limits direct metric evaluation for rare classes and requires application of patient-stratified validation and splitting schemes. At its current size, the dataset is not considered a complete training base for challenging multi-class diagnosis and is intended primarily for pilot analysis, external validation, domain shift analysis, interpretability studies, and subsequent expansion.
\item \textbf{Regional specificity.} The dataset was collected in Russian outpatient conditions and does not claim to represent the global population; external validation on independent samples from other regions is a direction for future work.
\item \textbf{Metadata completeness.} Some fields of the information model may be incomplete for individual records, which is typical for data collected in real clinical practice; such records are labeled and retained in the dataset with explicit indication of the completeness level.
\item \textbf{Verification level.} Records of benign and borderline neoplasms are confirmed by consensus of three specialists (\texttt{verification\_\allowbreak{}stage}~$= 2$), but lack histological confirmation, since biopsy is not clinically indicated in these cases. This limits the dataset's use as a ``gold standard'' for classifier specificity evaluation in benign classes; when strict verification is required, a subset should be selected using \texttt{verification\_\allowbreak{}stage}~$= 3$.
\end{enumerate}

\section{Conclusion}\label{sec:conclusion}

A three-component methodology for constructing a clinically verified dermoscopic image dataset has been developed, and a pilot dataset for medical informatics tasks has been created using this methodology.

\textbf{Main results and contributions:}
\begin{enumerate}
\item A three-component architecture for clinical dermoscopic dataset construction has been formalized, ensuring reproducibility of the acquisition procedure, verifiable completeness of structured metadata, and explicit accounting of diagnostic label confirmation level.
\item A standard operating procedure for mobile dermoscopic acquisition has been developed, regulating optical conditions, acquisition parameters, illumination conditions, and technical image quality control; the SOP is applicable across various clinical sites without specialized stationary equipment.
\item A 16-field metadata information model in ISIC-compatible notation has been developed, organized into six clinically oriented blocks and including a field for expert annotation of dermoscopic structures, which creates the basis for subsequent quantitative assessment of model interpretability.
\item A multi-stage procedure for expert verification of diagnostic labels has been formalized with explicit recording of the reliability level in the \texttt{verification\_\allowbreak{}stage} field, ensuring correct separation of clinical hypothesis from confirmed label.
\item Using the developed methodology, a clinically verified dataset has been constructed comprising 1026 unique dermoscopic images from 443 patients, 9 nosological categories, and histologically confirmed malignant classes; the dataset can be used as a clinically verified local resource for independent testing, interpretability analysis, domain adaptation, and subsequent expansion of training samples.
\end{enumerate}

The practical significance of the work lies in the formalization of a reproducible approach to preparing clinical dermoscopic datasets. The proposed information model and verification procedure can be applied when constructing new medical image collections, externally validating machine learning models, analyzing domain shift, methodologically demonstrating algorithms, and comparatively evaluating interpretability under conditions close to outpatient practice. Further development of the work is related to augmentation of rare diagnostic classes, expansion of the geographic data coverage, and quantitative assessment of model attention area agreement with expert-annotated dermoscopic structures based on the \texttt{dermoscopic\_\allowbreak{}structures} field.

\makefinish


\begin{thebibliography}{30}

\bibitem{ham10000} Tschandl~P., Rosendahl~C., Kittler~H. The HAM10000 dataset, a large collection of multi-source dermatoscopic images of common pigmented skin lesions~// Scientific Data.\,---\,2018.\,---\,Vol.\,5.\,---\,Article~180161. DOI:~10.1038/sdata.2018.161.

\bibitem{isic2017} Codella~N.\,C.\,F., Gutman~D., Celebi~M.\,E., Helba~B., Marchetti~M.\,A., Dusza~S.\,W., et~al. Skin lesion analysis toward melanoma detection: A challenge at the 2017 International Symposium on Biomedical Imaging~// Proceedings of the IEEE International Symposium on Biomedical Imaging.\,---\,2018.\,---\,P.\,168--172. DOI:~10.1109/ISBI.2018.8363547.

\bibitem{isic2020} Rotemberg~V., Kurtansky~N., Betz-Stablein~B., Caffery~L., Chousakos~E., Codella~N., et~al. A patient-centric dataset of images and metadata for identifying melanomas using clinical context~// Scientific Data.\,---\,2021.\,---\,Vol.\,8.\,---\,Article~34. DOI:~10.1038/s41597-021-00815-z.

\bibitem{bcn20000} Hernandez-Perez~C., Combalia~M., Podlipnik~S., Codella~N.\,C.\,F., Rotemberg~V., Halpern~A.\,C., et~al. BCN20000: Dermoscopic lesions in the wild~// Scientific Data.\,---\,2024.\,---\,Vol.\,11.\,---\,Article~641. DOI:~10.1038/s41597-024-03387-w.

\bibitem{padufes20} Pacheco~A.\,G.\,C., Lima~G.\,R., Salomao~A.\,S., Krohling~B., Biral~I.\,P., de~Angelo~G.\,G., et~al. PAD-UFES-20: A skin lesion dataset composed of patient data and clinical images collected from smartphones~// Data in Brief.\,---\,2020.\,---\,Vol.\,32.\,---\,Article~106221. DOI:~10.1016/j.dib.2020.106221.

\bibitem{daneshjou2022} Daneshjou~R., Vodrahalli~K., Novoa~R.\,A., Jenkins~M., Liang~W., Rotemberg~V., et~al. Disparities in dermatology AI performance on a diverse, curated clinical image set~// Science Advances.\,---\,2022.\,---\,Vol.\,8, no.~31.\,---\,Article~eabq6147. DOI:~10.1126/sciadv.abq6147.

\bibitem{groh2021} Groh~M., Harris~C., Soenksen~L.\,R., Lau~F., Han~R., Kim~A., Koochek~A., Badri~O. Evaluating deep neural networks trained on clinical images in dermatology with the Fitzpatrick 17k dataset~// Proceedings of the IEEE/CVF Conference on Computer Vision and Pattern Recognition Workshops.\,---\,2021.\,---\,P.\,1820--1828.

\bibitem{kinyanjui2020} Kinyanjui~N.\,M., Odonga~T., Cintas~C., Codella~N.\,C.\,F., Panda~R., Sattigeri~P., Varshney~K.\,R. Estimating skin tone and effects on classification performance in dermatology datasets~// Proceedings of the Machine Learning for Health NeurIPS Workshop.\,---\,2020.\,---\,P.\,18--30.

\bibitem{esteva2017} Esteva~A., Kuprel~B., Novoa~R.\,A., Ko~J., Swetter~S.\,M., Blau~H.\,M., Thrun~S. Dermatologist-level classification of skin cancer with deep neural networks~// Nature.\,---\,2017.\,---\,Vol.\,542.\,---\,P.\,115--118. DOI:~10.1038/nature21056.

\bibitem{wen2022} Wen~D., Khan~S.\,M., Xu~A.\,J., Ibrahim~H., Smith~L., Caballero~J., Zepeda~L., de~Blas~Perez~C., Denniston~A.\,K., Liu~X., Matin~R.\,N. Characteristics of publicly available skin cancer image datasets: A systematic review~// The Lancet Digital Health.\,---\,2022.\,---\,Vol.\,4, no.~1.\,---\,P.\,e64--e74. DOI:~10.1016/S2589-7500(21)00252-1.

\bibitem{brinker2019} Brinker~T.\,J., Hekler~A., Enk~A.\,H., Klode~J., Hauschild~A., Berking~C., Schilling~B., Haferkamp~S., Schadendorf~D., Holland-Letz~T., Utikal~J.\,S., von~Kalle~C. Deep neural networks are superior to dermatologists in melanoma image classification~// European Journal of Cancer.\,---\,2019.\,---\,Vol.\,119.\,---\,P.\,11--17. DOI:~10.1016/j.ejca.2019.05.023.

\bibitem{ph2} Mendonca~T., Ferreira~P.\,M., Marques~J.\,S., Marcal~A.\,R.\,S., Rozeira~J. PH2\,---\,A dermoscopic image database for research and benchmarking~// 2013 35th Annual International Conference of the IEEE Engineering in Medicine and Biology Society (EMBC).\,---\,2013.\,---\,P.\,5437--5440. DOI:~10.1109/EMBC.2013.6610779.

\bibitem{combalia2022} Combalia~M., Codella~N., Rotemberg~V., Carrera~C., Dusza~S., Gutman~D., et~al. Validation of artificial intelligence prediction models for skin cancer diagnosis using dermoscopy images: The 2019 International Skin Imaging Collaboration grand challenge~// The Lancet Digital Health.\,---\,2022.\,---\,Vol.\,4, no.~5.\,---\,P.\,e330--e339. DOI:~10.1016/S2589-7500(22)00021-8.

\bibitem{daneshjou2021} Daneshjou~R., Smith~M.\,P., Sun~M.\,D., Rotemberg~V., Zou~J. Lack of transparency and potential bias in artificial intelligence data sets and algorithms: A scoping review~// JAMA Dermatology.\,---\,2021.\,---\,Vol.\,157, no.~11.\,---\,P.\,1362--1369. DOI:~10.1001/jamadermatol.2021.3129.

\bibitem{gebru2021} Gebru~T., Morgenstern~J., Vecchione~B., Vaughan~J.\,W., Wallach~H., Daume~H. III, Crawford~K. Datasheets for datasets~// Communications of the ACM.\,---\,2021.\,---\,Vol.\,64, no.~12.\,---\,P.\,86--92. DOI:~10.1145/3458723.

\bibitem{winkler2019} Winkler~J.\,K., Fink~C., Toberer~F., Enk~A., Deinlein~T., Hofmann-Wellenhof~R., Thomas~L., Lallas~A., Blum~A., Stolz~W., Haenssle~H.\,A. Association between surgical skin markings in dermoscopic images and diagnostic performance of a deep learning convolutional neural network for melanoma recognition~// JAMA Dermatology.\,---\,2019.\,---\,Vol.\,155, no.~10.\,---\,P.\,1135--1141. DOI:~10.1001/jamadermatol.2019.1735.

\bibitem{tschandl2020} Tschandl~P., Rinner~C., Apalla~Z., Argenziano~G., Codella~N., Halpern~A., et~al. Human-computer collaboration for skin cancer recognition~// Nature Medicine.\,---\,2020.\,---\,Vol.\,26, no.~8.\,---\,P.\,1229--1234. DOI:~10.1038/s41591-020-0942-0.

\bibitem{kozachok2025dataset} Kozachok~E.\,S. Dermoscopic image dataset with high-quality annotation of clinically significant features for the diagnosis of melanocytic neoplasms~// Izvestiya YuZGU. Seriya: Upravleniye, vychislitel'naya tekhnika, informatika. Meditsinskoe priborostroyeniye [Proceedings of SWSU. Series: Control, Computer Engineering, Information Science. Medical Instrumentation].\,---\,2025.\,---\,Vol.\,15, No.\,3.\,---\,P.\,93--111. DOI:~10.21869/2223-1536-2025-15-3-93-111. (In Russian)

\bibitem{kozachok2025screening} Kozachok~E.\,S., Seregin~S.\,S., Kozachok~A.\,V., Eletskiy~K.\,V., Samovarov~O.\,I. Screening methodology for early differential diagnosis of skin neoplasms using mobile dermoscopy~// Vrach i informatsionnyye tekhnologii [Doctor and Information Technologies].\,---\,2025.\,---\,No.\,3.\,---\,P.\,50--64. DOI:~10.25881/18110193\_2025\_3\_50. (In Russian)

\bibitem{kozachok2025sppvr1} Kozachok~E.\,S., Seregin~S.\,S., Kozachok~A.\,V., Eletskiy~K.\,V., Samovarov~O.\,I. Intelligent clinical decision support system for skin neoplasm diagnosis based on dermoscopic image analysis~// Izvestiya YuZGU. Seriya: Upravleniye, vychislitel'naya tekhnika, informatika. Meditsinskoe priborostroyeniye [Proceedings of SWSU. Series: Control, Computer Engineering, Information Science. Medical Instrumentation].\,---\,2025.\,---\,Vol.\,15, No.\,3.\,---\,P.\,50--65. DOI:~10.21869/2223-1536-2025-15-3-50-65. (In Russian)

\bibitem{kozachok2025sppvr2} Kozachok~E.\,S., Seregin~S.\,S. Intelligent clinical decision support system for skin neoplasm diagnosis based on mobile dermoscopy~// Rossiyskiy zhurnal telemitsiny i elektronnogo zdravookhraneniya [Russian Journal of Telemedicine and E-Health].\,---\,2025.\,---\,Vol.\,11, No.\,3.\,---\,P.\,38--44. DOI:~10.29188/2712-9217-2025-11-3-38-44. (In Russian)

\end{thebibliography}
\end{document}